\documentclass[%
 reprint,
showpacs,preprintnumbers,
 amsmath,amssymb,
 aps,
pra,
]{revtex4-1}

\usepackage{graphicx}
\usepackage{dcolumn}
\usepackage{bm}
\usepackage{braket}

\usepackage{hyperref}

\usepackage{amssymb}

\begin{document}


\title{ The properties of squeezed optical states created in lossy cavities}

\author{Hossein Seifoory}
\email{hossein.seifoory@queensu.ca}
\author{Sean Doutre}
\author{Marc. M. Dignam}
 \affiliation{Department of Physics, Engineering Physics and Astronomy, Queen's University, Kingston, Ontario K7L 3N6, Canada}

\author{J. E. Sipe}%
\affiliation{Department of Physics and Institute for Optical Sciences, University of Toronto, 60 St. George Street, Toronto, Ontario M5S 1A7, Canada}%

\date{\today}
\begin{abstract}
We investigate theoretically the properties of squeezed states generated using degenerate parametric down conversion in lossy cavities. We show that the Lindblad master equation, which governs the evolution of this system, has as its solution a squeezed thermal state with an effective temperature and squeezing parameter that depends on time. We derive analytical solutions for the time-evolution of quadrature noise, thermal photon number, squeezing parameter, and total photon number under different pumping regimes. We also find the steady state limits of the quadrature noises and discuss the $ g^{(2)} $ factor of the generated light inside the cavity in the steady state.  
\end{abstract}

\maketitle


\section{\label{sec:level1}Introduction}\
The theory of field quantization has its origins in Dirac's work of 1927~\cite{Dirac243,fox2006quantum}.
One of the most important consequences of quantizing the electromagnetic field is the existence of vacuum fluctuations. In any state of the field, even the vacuum, the electric and magnetic fields at any given time deviate from their mean. These fluctuations are responsible for phenomena such as spontaneous emission~\cite{spontaneous}, zero-point energy~\cite{zero_point_energy}, Lamb shift~\cite{Lamb1,Lamb2}, quantum beats~\cite{QuantumBeat} and the Casimir effect~\cite{Casimir1, Casimir2}. More directly, the noise resulting from these fluctuations sets a limit on the accuracy of interferometric measurements. Squeezed states suppress this noise in a particular quadrature, and can be used to measure extremely weak signals, such as found in gravity-wave interferometers~\cite{LIGO1, LIGO2, LIGO3}.\par
Squeezed states can be obtained using nonlinear optical processes such as four-wave mixing~\cite{four-wave} and parametric down conversion~\cite{parametric_down_conversion}, or from resonant fluorescence of coherently driven single-photon emitters~\cite{reduced_quantum,PhysRevLett.59.198, PhysRevLett.81.3635, PhysRevLett.109.013601}, photonic crystal microcavities~\cite{Banaee:08}, and even from an individual atom~\cite{one_atom}. \par
Kim \textit{et al.}~\cite{PhysRevA.40.2494} have studied the statistical properties of the squeezed thermal states, including the photon number distribution, quasiprobability function, and the second order correlation function. They found, for example, that in the weak squeezing limit these state exhibit strong photon-bunching~\cite{PhysRevA.40.2494}. It has been shown that the phase sensitivity of Mach-Zehnder interferometry can be enhanced using single-mode squeezed thermal light~\cite{PhysRevA.89.053822}, and that when a squeezed thermal field is used as a light source in ghost imaging the visibility will greatly improve~\cite{PhysRevA.93.013822}. Moreover, a large number of phenomena have been predicted by studying the interaction of the squeezed field with atomic systems~\cite{PhysRevLett.58.2539, PhysRevA.40.3796, 0295-5075-24-6-005, PhysRevA.43.6247, PhysRevA.43.6258, PhysRevA.45.2025}, which was first investigated by Milburn~\cite{two-levelandsqueezedlight}. It has also been shown that the squeezed fields can be employed in some spectroscopic methods to improve their sensitivity~\cite{PhysRevLett.75.3426, PhysRevLett.68.3020, ref1}. In order to assess the properties and suitability of the light generated in a leaky cavity for the above-mentioned applications, it is important to quantify the dependence of the photon noise and degree of anti-bunching on the  pump and loss rates in a pumped, leaky cavity. \par 
In this paper we investigate a leaky cavity in which a classical pump field is used to generate a squeezed state via parametric down conversion. We limit ourselves to only one leaky signal mode and pump this mode at twice its natural frequency. We derive time-dependent and steady-state solutions for the of quadrature noise, thermal photon number, squeezing parameter, total photon number and second order correlation function under different pumping regimes.\par
The paper is organized as follows. In section II, we present the model of the dynamics of the system using Lindblad master equation. In section III, we then develop a solution to the Lindblad master equation using density matrix method, which reduces the problem to solving three coupled first-order differential equations for the squeezing amplitude, squeezing angle and average thermal photon number. Next, in Section IV, we review some properties of squeezed and squeezed thermal states and in Section V, we present the results in the weak and strong pumping regime for squeezed states generated in lossy cavities. Finally, we present our conclusions in Section VI.
\section{\label{sec:level1}the Lindblad master equation}
There are many ways to achieve squeezed states. Here we focus on degenerate parametric down conversion using a material with a nonlinear $ \chi^{(2)} $ response. Briefly, in this method a single pump photon of frequency $ \omega_P $ is converted into two signal photons of frequency $ \omega=\omega_P /2 $. The Hamiltonian takes the form
\begin{equation}
{H}=\hbar \omega {b}^{\dagger} {b}+\hbar \omega_P {a}^{\dagger} {a}+i\hbar \chi_{eff}^{(2)}({b}^{2}{a}^{\dagger}-{b}^{\dagger 2} {a}),
\end{equation}
where $ a $ and $ b $ are annihilation operators of photons in the pump and signal modes respectively, and $ \chi_{eff}^{(2)} $ is the effective second order nonlinear susceptibility.
In the parametric approximation, the input photons are assumed to come from a strong pump which we approximate as a time-dependent coherent state $\ket{\alpha(t)}$. This leads to the following Hamiltonian for the signal photons
\begin{equation}
{H}=H_0+ (\alpha(t) \gamma {b}^{\dagger} {b}^{\dagger}+\alpha^*(t) \gamma^* {b} {b}), 
\end{equation}
where $H_0= \hbar \omega {b}^{\dagger} {b} $ and $ \gamma $ is the complex coupling between the pump and signal modes, which depends on the material properties of the system and the pump-signal mode overlap. 
The cavity has a resonance at $ \omega $, with a quality factor, $Q=\omega/\Gamma$, where $ \Gamma $ is the intensity decay constant of the mode. 
The dynamics of this system can be modelled using the Lindblad master equation
\begin{equation}\label{lindblad01}
{\dot{\rho}}=-\frac{i}{\hbar} [{H},{\rho}] +\Gamma ({b}{\rho}{b}^{\dagger}-\frac{1}{2}{b}^\dagger{b}{\rho}-\frac{1}{2}{\rho}{b}^{\dagger}{b}),
\end{equation} 
where $ {\rho} $ is the density operator. If the pump is on-resonance with the cavity, then $ \alpha(t)= \alpha_0 e^{-2i\omega t} $, and in the interaction picture
\begin{equation}
{U_{int}}(t)=e^{ -it (\gamma^* \alpha_{0}^* {b} {b} + \gamma \alpha_{0} {b}^\dagger {b}^\dagger)/\hbar}.
\end{equation}
is the time evolution operator with the neglect of loss. This has the form of squeeze operator of $ {S}(\xi) =\exp[\frac{1}{2}(\xi^{*}{b}^{2}-\xi {b}^{\dagger 2})] $ for $ \xi=2i \gamma \alpha_0 t /\hbar $. When acting on the vacuum, the squeeze operator gives
\begin{equation}
{S}(\xi)\ket{0}=\ket{\xi},
\end{equation}
where $ \ket{\xi}$ is a squeezed vacuum state (SVS). The squeezing parameter is generally complex, and we write it in the form $ \xi = u e^{i\phi} $.\par
In the presence of loss, the solution is considerably more complicated. One can calculate the time evolution of the density matrix numerically. Before performing such a numerical solution, it is advantageous to introduce the dimensionless parameter $ \wp \equiv (4\alpha_{0}\gamma)/(\hbar \Gamma) $, which  represents the ratio of the pumping strength to the loss rate. Then, Eq.~(\ref{lindblad01}) can be rewritten in the interaction picture as
\begin{equation} \label{lindblad02}
\frac{{\dot{\rho}}}{\Gamma}= -\frac{i}{\hbar} [H',{\rho} ]+({b}{\rho}{b}^{\dagger}-\frac{1}{2}{b}^\dagger{b}{\rho}-\frac{1}{2}{\rho}{b}^{\dagger}{b}),
\end{equation}
where $H'=(\wp^{*}\hbar{b}^{2}+\wp\hbar{b}^{\dagger 2})/4  $, such that the only two parameters that govern the system are $ \Gamma t $ and $ \wp $.\par   
The numerical solution of the Lindblad master equation and consequently the expectation values of the quadrature operators can be obtained by employing a basis of number states, as long as the photon number does not become so large that an impractically large basis is required~\cite{Qutip}. In this work, however, we take a different approach. In the following section, we show that rather than resorting to a numerical solution to the master equation, the exact solution for the density operator takes the form of a squeezed thermal state (STS), where the squeezing parameter and effective temperature are determined by a simple set of coupled first order differential equations.

\section{\label{sec:level2}analytic solution of the lindblad master equation}
As we show in this section, the solution of Eq.~(\ref{lindblad01}) can be written exactly as
        \begin{equation}\label{general_solution}
        {\rho} (t)={S}(\xi(t)){\rho_T}(\beta(t)) {S^{\dagger}}(\xi(t)),
        \end{equation}
where $ S(\xi(t)) $ is the squeezing operator, $ \xi(t) $ depends on time and
\begin{equation}
{\rho}_T(\beta(t))=(1-e^{-\beta (t)\hbar \omega})e^{-\beta (t) H_0}
\end{equation}
is a density operator describing a thermal state at an (effective) time-dependent temperature $ k_B T(t)=1/\beta (t) $. This temperature should of course be distinguished from the temperature of the reservoir, which by assumption is zero. Thus the evolving state is a squeezed thermal state. From the form of Eq.~(\ref{general_solution}), we see that one can use this solution to describe the evolution of a variety of initial states, including a vacuum state, thermal state, squeezed state or squeezed thermal state by simply changing the initial conditions.

In describing the solution, rather than parameterizing the thermal state using $ \beta(t) $ it is easier to use
\begin{equation}\label{n and beta}
n_{th}(t)\equiv \frac{1}{e^{\beta (t) \hbar \omega} - 1},
\end{equation}  
which is the expectation value of the photon number in the thermal state. Although we could directly confirm our exact solution by plugging Eq.~(\ref{general_solution}) into Eq.~(\ref{lindblad01}), here we derive it by a method that may be useful in related systems when we cannot find an exact solution but must search for an approximate one. We seek a solution of Eq.~(\ref{lindblad01}) of the form
        \begin{equation}\label{general_solution1}
        {\rho} (t)={S}(\xi(t)){\rho}_T^{1/2} (n_{th}) {O}(t) {\rho}_T ^{1/2}(n_{th}) {S^{\dagger}}(\xi(t)),
        \end{equation}
where the initial value of $ n_{th}(t=0)=0 $ is chosen when our initial state is the vacuum. Comparing Eq.~(\ref{general_solution}) with Eq.~(\ref{general_solution1}), it can be shown that $ {O} $ should be the identity operator for all times, $  O (t)=I $. 
From Eq.~(\ref{general_solution1}), it is easily seen that
\begin{equation}
O(t)=\rho^{-1/2}_T S^{\dagger}(\xi)\rho (t) S(\xi) \rho^{-1/2}_T .
\end{equation} 
As we show in the Appendix, forcing $ O(t) $ to be identity operator for all times leads to the following three coupled differential equations for squeezing amplitude, squeezing phase and average thermal photon number:
\begin{equation}\label{u}
\frac{du(t)}{dt}=\frac{i}{\hbar}(\gamma \alpha (t) e^{-i \phi(t)}-\gamma ^{*} \alpha ^{*} (t) e^{i \phi(t)})-\frac{\Gamma cs}{2n_{th}(t)+1},
\end{equation}
\begin{equation}\label{phi}
\frac{d\phi(t)}{dt}=-2\omega+\frac{1}{\hbar}\frac{c^2 + s^2}{cs}(\gamma \alpha (t) e^{-i \phi(t)}+\gamma ^{*} \alpha ^{*} (t) e^{i \phi(t)}),
\end{equation}
\begin{equation}\label{n}
\frac{dn_{th}(t)}{dt}=\Gamma (s^2 - n_{th}(t)),
\end{equation}
where $ s \equiv \sinh u(t) $ and $ c \equiv \cosh u(t) $. The last equation shows that the state gains thermal photons at a rate given by $ \Gamma \sinh^2 u(t) $ and loses these photons at a rate $ \Gamma n_{th}(t) $. If the system is initially prepared as a pure SVS and if there is no leakage ($ \Gamma=0 $), then $ \sinh^2 u(t) $ is simply the number of photons in the system. We later show that in a STS, $ \sinh^2 u(t) $ cannot simply be interpreted as the number of photons produced by the action of the squeeze operator.\par
In all that follows, we turn the field on at $ t=0 $ ($ \alpha (t)=0 $ for $ t<0 $) and we take the initial state to be the vacuum state, so that $ u(0)=0 $.  Thus, because $ \sinh u(0) = 0 $, the right hand side of Eq.~(\ref{phi}) is only finite if at $ t=0 $ ($ \alpha (t)=0 $ for $ t<0 $ and $ \alpha (t)=\alpha_0 e^{-2i\omega t} $ for $ t>0 $)
\begin{equation}\label{condition}
(\gamma \alpha(t) e^{-i \phi(t)}+\gamma ^{*} \alpha(t) ^{*} e^{i \phi(t)})=0.
\end{equation}
We obtain our solution by enforcing this condition for all times. We then see that the phase then evolves according to
\begin{equation}\label{minus2omega}
\frac{d\phi(t)}{dt}=-2\omega,
\end{equation}
which is trivially solved as $ \phi(t)= -2\omega t+\phi_0 $. It then follows from Eq.~(\ref{condition}) that for $ t\geq 0 $ we must have  
\begin{equation}\label{gammaAlpha}
\gamma \alpha(t)=-i \eta e^{-i(2 \omega t-\phi_0)},
\end{equation} 
where $ \eta $ is a real constant. We thus only have to solve Eqs.~(\ref{u}) and (\ref{n}) with the conditions given by Eqs.~(\ref{minus2omega}) and (\ref{gammaAlpha}). Following these conditions, Eq.~(\ref{u}) can be rewritten as 
\begin{equation}\label{modified_u}
\frac{1}{\Gamma}\frac{du(t)}{dt}=\frac{g}{2}-\frac{cs}{2n_{th}(t)+1},
\end{equation}
where $ g $ is a \textit{real} constant given by $ g=i\wp $, which without loss of generality can be taken as positive, and the time dependence of $ n_{th}$ is given by Eq.~(\ref{n}) \cite{extra1}. 
Before presenting the results for our dynamical equations, we first examine some of the properties of squeezed thermal states. In all that follows, we consider that the pump phase is chosen such that $ \phi_0 = 0 $.  This is not restrictive, because it is easy to show that choosing a different pump phase only results in a rotation in the quadrature plane and does not change the physical content of the results. 
\section{\label{sec:level3}Squeezing and some Properties of The Squeezed Thermal States}
To quantify the vacuum fluctuations, we define the annihilation operator in terms of Hermitian quadrature operators, $ X $ and $ Y$, as
\begin{equation} \label{X_and_Y}
b=e^{i\phi(t)/2}(\frac{X+iY}{2}).
\end{equation}
With this definition of the quadrature operators, we remove the trivial free evolution in the usual way~\cite{Squeezed_state_with_thermal_noise} so that we can focus on the effects of pumping and loss.\par 
The RMS deviations for the quadratures obey the uncertainty relation
\begin{equation}
\Delta{X} \Delta{Y} \geq 1.
\end{equation} 
Both vacuum and coherent states are minimum uncertainty states where the noise is evenly distributed between $ \Delta X $ and $ \Delta Y $. It is possible to reduce the noise in one quadrature by increasing the noise in the other. Such squeezing can lead to a reduction in the uncertainty in the amplitude of the electric field or in its phase.\par
The mean photon number of SVS is found to be
\begin{equation}\label{SVS_n}
\braket{n}=\braket{b^{\dagger}b} = \sinh^2 (u) .
\end{equation}
The squeezed thermal state is the Bose-Einstein weighted sum of the squeezed number state. The density matrix operator of a single-mode squeezed thermal state is
\begin{equation}\label{eq:rho n}
\rho = \sum_{m=0}^{\infty} \frac{n_{th}^{m}}{(n_{th}+1)^{m+1}}S(\xi)\ket{m}\bra{m}S^{\dagger}(\xi),
\end{equation}
where $ n_{th} $ is the thermal photon number. The variances of quadrature operators and the number operator for the squeezed thermal state are~\cite{PhysRevA.40.2494}
\begin{equation} \label{eq:delta x}
\braket{(\Delta X)^2}=(2n_{th}+1)e^{-2u},
\end{equation}
\begin{equation}\label{eq:delta y}
\braket{(\Delta Y)^2}=(2n_{th}+1)e^{2u},
\end{equation}
and
\begin{equation}\label{eq:b bdag}
\braket{n}=\braket{b^{\dagger}b}=n_{th} \cosh(2u)+\sinh^2 (u).
\end{equation} 
Moreover, the correlation between photons can be determined by the normalized single-mode second order correlation function using creation and annihilation operators as
\begin{equation}
g^{(2)}=\frac{\braket{b^{\dagger} b^{\dagger} b b}}{\braket{b^{\dagger} b}^2}.
\end{equation}
Using Eqs.~(\ref{eq:rho n}) and (\ref{eq:b bdag}), the second order correlation function can be written in terms of thermal photon number and squeezing parameter as~\cite{PhysRevA.40.2494}
\begin{equation}\label{eq:g2_factor}
g^{(2)}=2+\frac{(n_{th}+\frac{1}{2})^2 \sinh^2 (2u)}{(n_{th} \cosh (2u)+\sinh^2 (u))^2},
\end{equation}
which quantifies the intensity fluctuations in a classical picture and photon bunching in a quantum picture.

\section{\label{sec:level4}results}
\begin{figure}
         
                \includegraphics[scale=0.45]{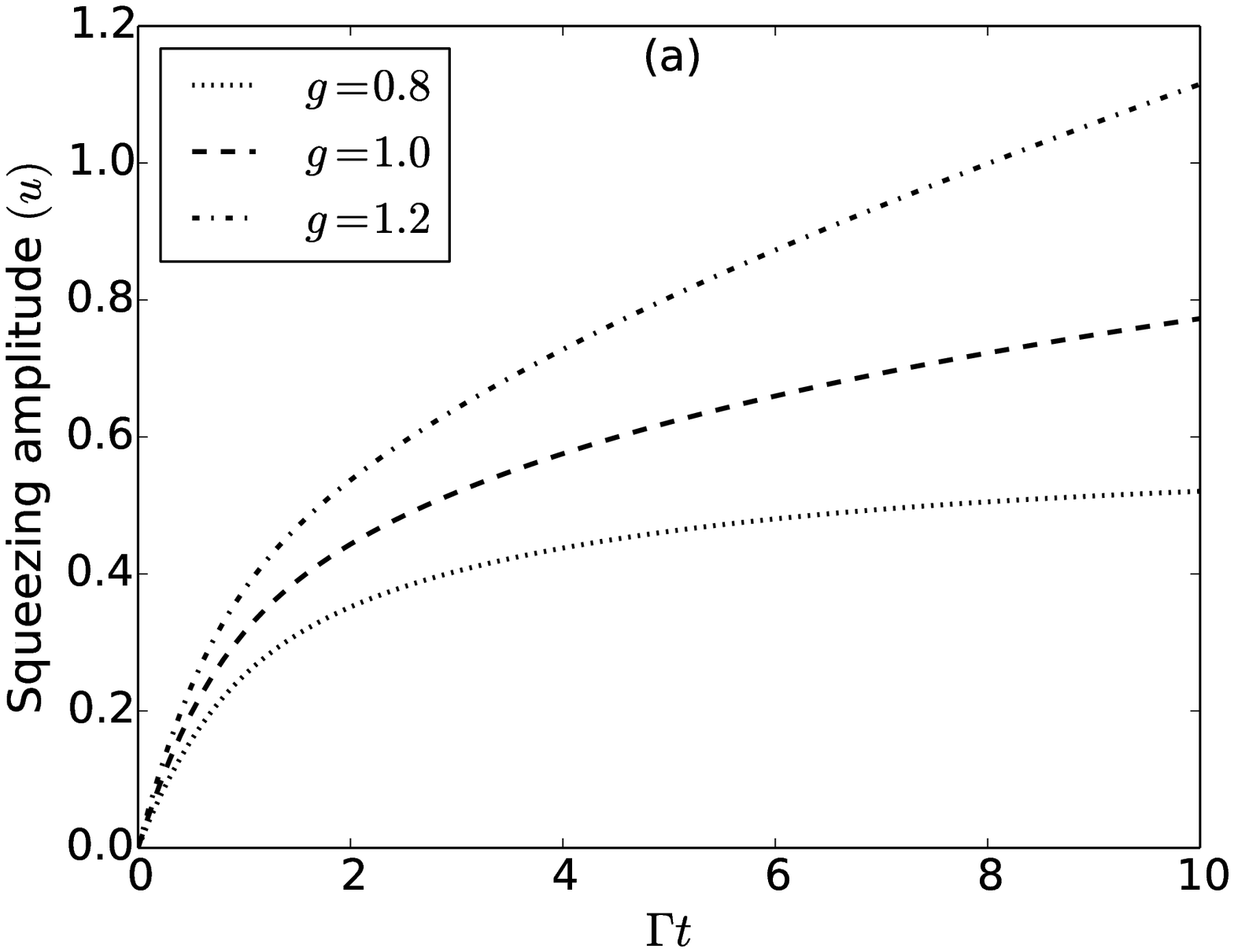}
                        \includegraphics[scale=0.45]{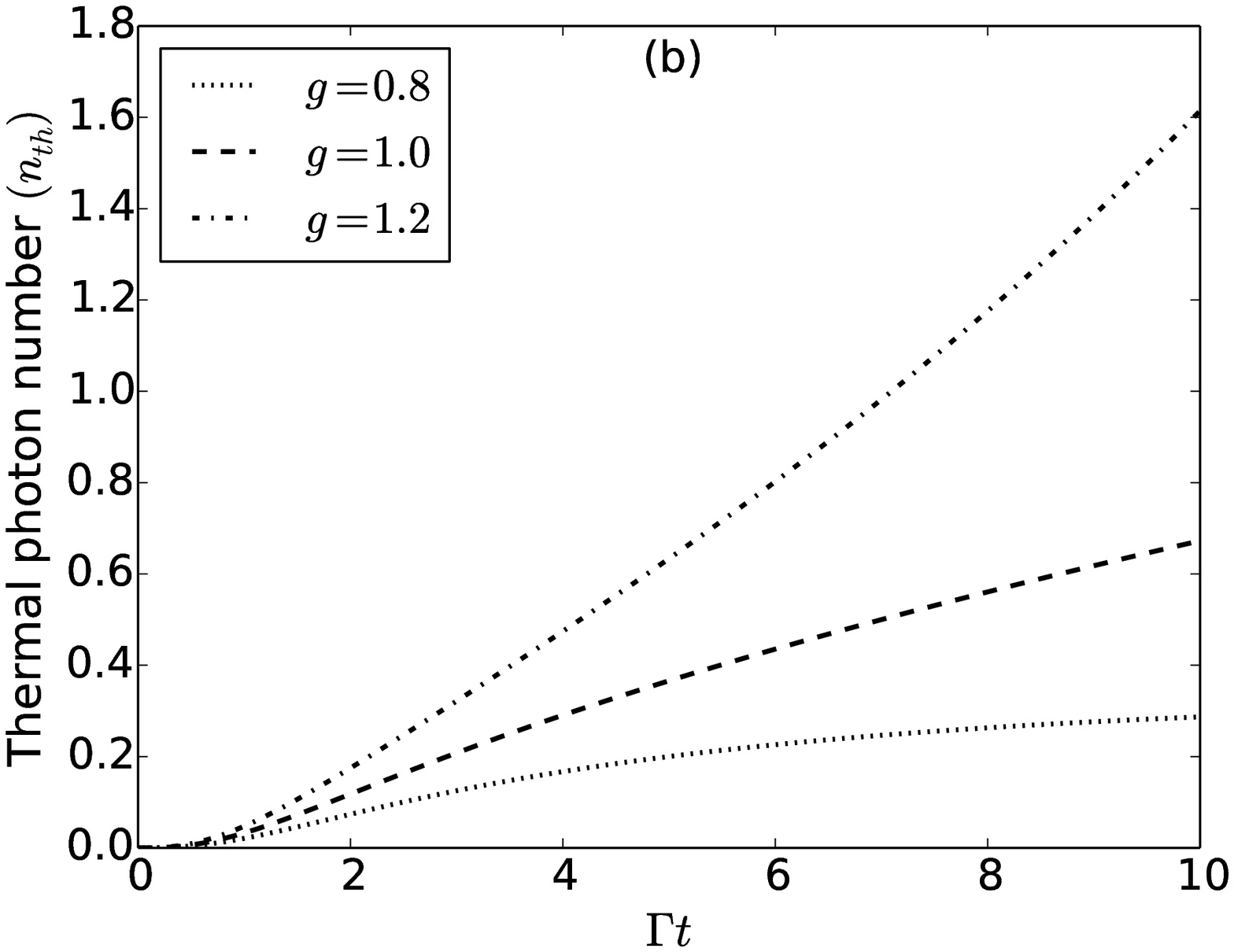}
                      
                \caption{The time dependence of (a) the squeezing amplitude and (b) the thermal photon number in the case of weak pumping ($ g=0.8 $), critical pumping ($ g=1.0 $) and strong pumping ($ g=1.2 $).} 
                \label{fig:m_ODE}
                
                \end{figure}

\begin{figure}          
                \includegraphics[scale=0.45]{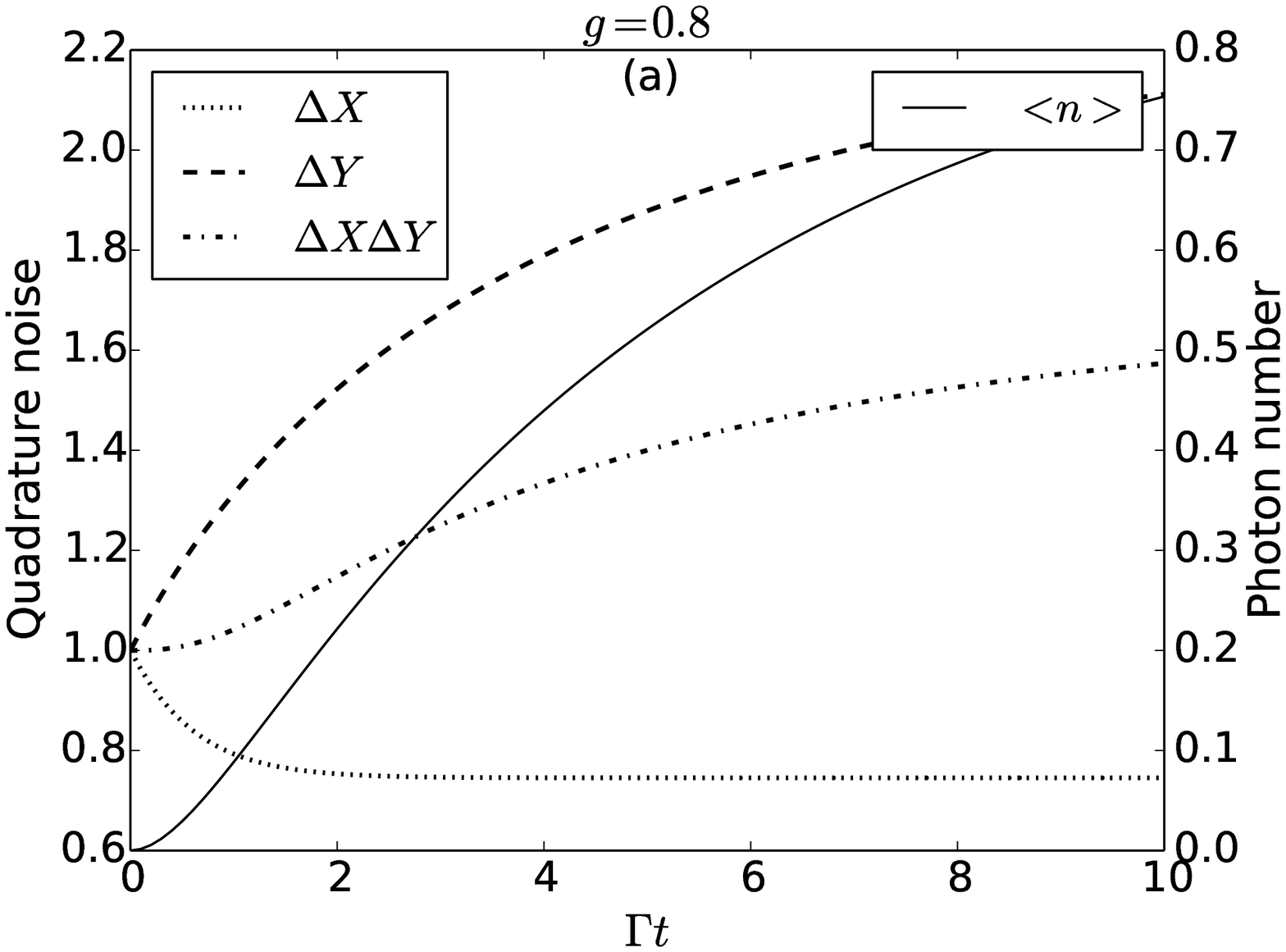}
                        \includegraphics[scale=0.45]{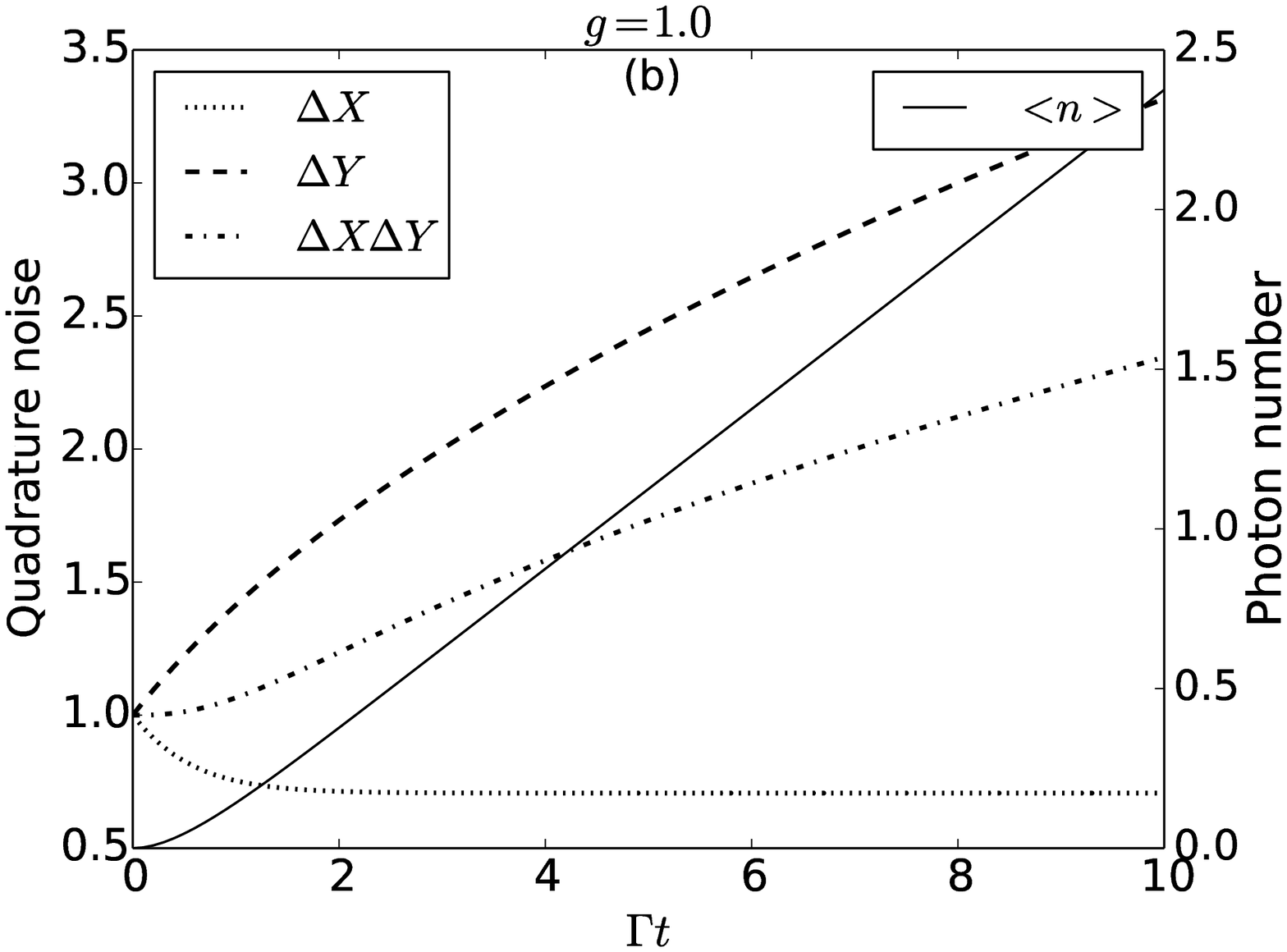}
                                \includegraphics[scale=0.45]{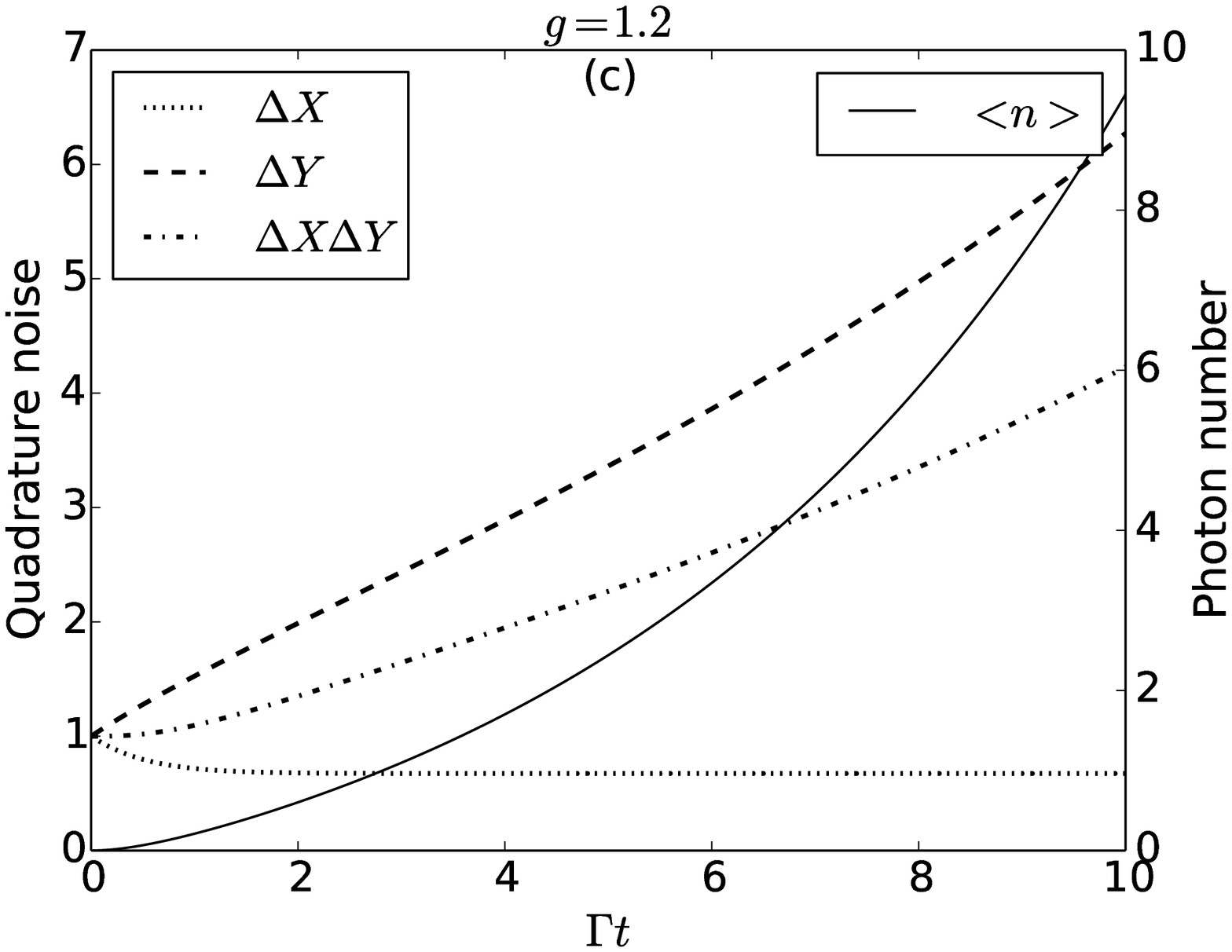}
                                
                \caption{The quadrature noises and the expectation value of the total photon number as a function of time for (a) weak pumping ($ g=0.8 $), (b) critical pumping ($ g=1.0 $) and (c) strong pumping ($ g=1.2 $).}
                \label{fig:ODE_noise}
                \end{figure}

The two coupled Eqs.~(\ref{n}) and (\ref{modified_u}) were solved numerically in three different pumping regimes, weak, critical, and strong. The dynamics of the thermal photon number and the squeezing amplitude are plotted in Fig.~\ref{fig:m_ODE} for different pumping regimes. The parameter, $g$, quantifies the pumping level, with the critical pumping value, $g=1$, marking the dividing line between weak and strong pumping. In the weak pumping regime ($g<1$), both the thermal photon number and the squeezing amplitude saturate and reach a steady state value.  In contrast, for strong pumping ($g>1$), both of these quantities increase rapidly, with the thermal photon number increasing exponentially at long times. 
Using the time dependence of $n_{th}$ and $u$ and employing Eqs.~(\ref{eq:delta x}), (\ref{eq:delta y}) and (\ref{eq:b bdag}), one can calculation the time evolution of the quadrature noises and the photon number expectation value. In Fig.~\ref{fig:ODE_noise} we plot the total photon number, the quadrature noises and the product of quadrature noises ($ \Delta X \Delta Y
$) as they evolve in time. Note that for a pure squeezed state, $ \Delta X \Delta Y=1$, so the deviation from 1 is a measure of departure from a pure squeezed state. \par
We consider first the expectation value of the total photon number. As expected, this increases as $ g $ is increased. In the weak pumping regime, we observe that $ \braket{n} $ approaches a steady state value, while in the strong pumping
regime $ \braket{n} $ increases approximately exponentially at later times. We note that in all cases,
at early times ($ \Gamma t<1 $) the average photon number is given approximately by $ \sinh^2 u(t) $, which is the number of photons in a pure squeezed vacuum state.
At longer times, we find that $ \braket{n} $ is not simply equal to $ n_{th}(t)+\sinh^2 u(t)
$ (the addition of the thermal and "squeezed" photons) and that in the strong pumping regime, the deviation from this increases rapidly with time.

Now we turn to the quadrature uncertainties. As the value of $g$ is increased, we see that $ \Delta X$ decreases, as expected.  For the range of $g$ considered, the squeezing of $ \Delta X $ is rather modest, only reaching a value of $ 0.67 $ for $g=1.2$. Of course, if we increase $g$ further, then much stronger squeezing can be achieved. However, the price that one pays for the increase in squeezing in $X$ is an increase in $ \Delta Y$. In contrast
to the case of a pure squeezed state, the product of the two quadrature uncertainties
does not simply equal 1, except at $t=0$. From Eqs. (\ref{eq:delta x}) and (\ref{eq:delta y}), we see that $ \Delta X \Delta Y =(2n_{th} + 1)$, so that this product increases in direct proportion to the number of thermal photons. 
We will consider this tradeoff in more detail when we consider the strong
pumping regime below.\par

We close this section by noting that the results we have presented agree to within numerical precision with those obtained using a direct numerical solution to the master equation as implemented in QuTiP~\cite{Qutip}, thus verifying our analytic solution.\par
\subsection{Weak pumping regime}
\begin{figure}          
                \includegraphics[scale=0.45]{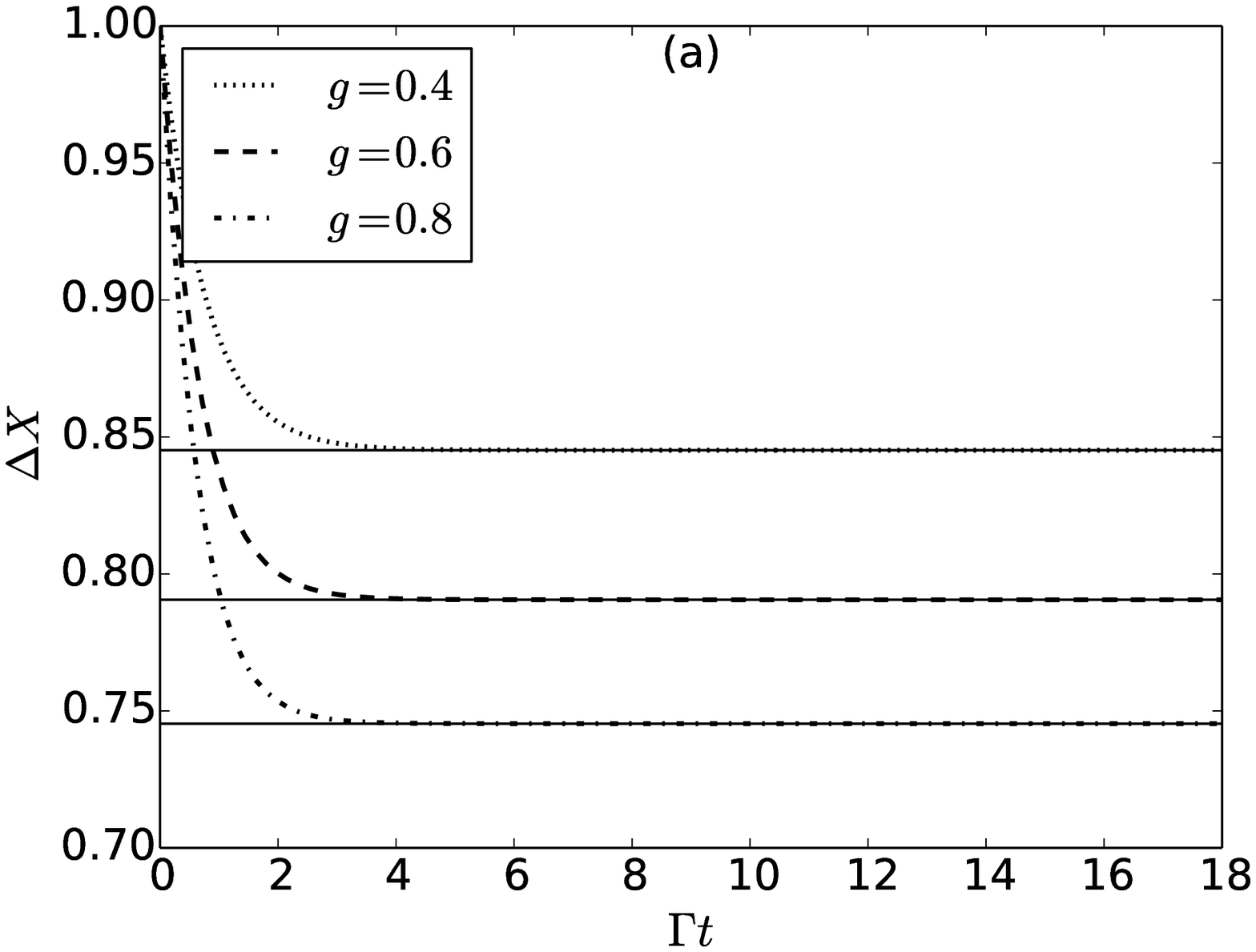}
                \includegraphics[scale=0.45]{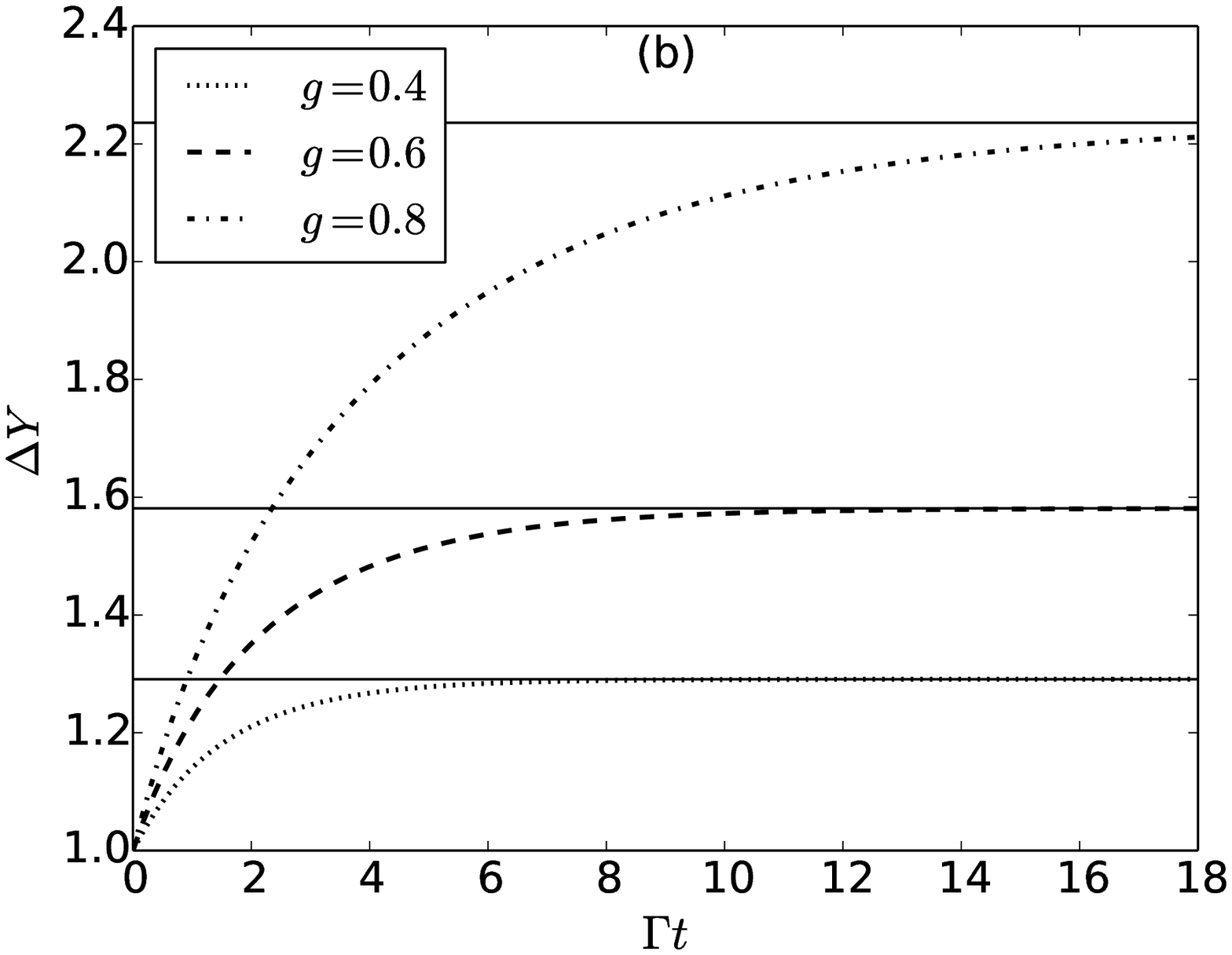}
              
                \caption{The quadrature noises in (a) $X$ and (b) $Y$ as a function of time for the sub-critical pump ratios of $ g=0.4, 0.6$ and $0.8 $. Also shown are the steady state limits (solid lines) given by Eqs.~(\ref{Xss}) and (\ref{Yss}).}
                \label{fig:limits Dx Dy}
                \end{figure}

In this section, we consider the weak pumping regime ($g<1$), with a focus on the steady state results. The steady state solution is obtained by setting the time derivatives for $ u $ and $ n_{th} $ (Eqs.~(\ref{n}) and (\ref{modified_u})) to zero, which gives
\begin{equation}\label{sinh_2}
\sinh^2 (u)=n_{th}
\end{equation}
and
\begin{equation}\label{tanh_2u}
g=\tanh (2u).
\end{equation}
Using Eqs.~(\ref{eq:delta x}) and (\ref{eq:delta y}) the variances of the quadrature operators in the steady state can be shown to be
\begin{equation}\label{Xss}
\Delta X_{ss} = \frac{1}{\sqrt{1+g}}
\end{equation} 
for any $g$ and
\begin{equation}\label{Yss}
\Delta Y_{ss} = \frac{1}{\sqrt{1-g}}
\end{equation}
when $g<1$. The condition for steady state operation can only be satisfied for weak pumping ($ g < 1 $).  
In Fig. ~\ref{fig:limits Dx Dy} we plot $ \Delta X $ (a) and $ \Delta Y $ (b) as a function of time for three different sub-critical pump ratios. In all cases we see that $ \Delta X $ decreases rapidly over a time of approximately $ \Gamma t=2 $ and then levels off, asymptotically approaching the steady state value given by Eq.~(\ref{Xss}). The presence of loss sets a limit on the noise suppression that can be achieved for a given pumping strength which is given by $ \Delta X > 1 / \sqrt{2}$ in the weak pumping regime. Note that this is the same limit obtained for light inside the cavity of an optical parametric oscillator \cite{garrison2014quantum,PhysRevA.30.1386}, as expected.\par
As seen in Fig.~\ref{fig:limits Dx Dy}(b), the stretched quadrature noise $ \Delta Y $ increases over a longer time ($ \Gamma t>10 $) before approaching the steady state value given by Eq.~(\ref{Yss}).
As the pump power is increased, a longer time is needed to reach the steady state; we will exploit this property in the strong pumping regime discussed in the next section. Note that the product of the uncertainties is given simply by
\begin{equation}
\Delta X_{ss} \Delta Y_{ss}= \frac{1}{\sqrt{1-g^2}},
\end{equation}
which diverges as $g$ approaches 1. This all indicates that if one is interested
in achieving strong squeezing in one quadrature without excessive stretching in the other, one should not operate in the weak pumping regime.\par
\begin{figure}          
                \includegraphics[scale=0.45]{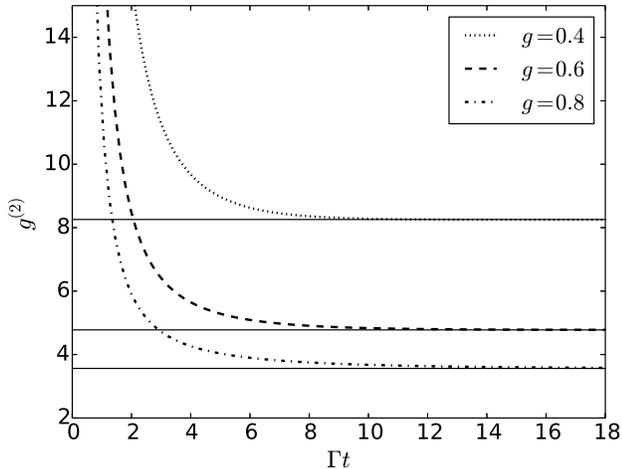}
           
                \caption{The second order correlation function as a function
of time for the sub-critical pumping values $ g=0.4, 0.6$ and $0.8 $. Also
shown (solid lines) are the steady state limits given by Eq.~(\ref{g2_steady}).}
                \label{fig:g2_factor}
                \end{figure}

We now consider the time-dependent and steady state behaviour of the
second order correlation function. The time dependence is plotted in Fig.~\ref{fig:g2_factor}
for three different sub-critical pumping powers, where we see that $ g^{(2)}
$ reaches steady state values that are greater than 3 in all cases. It is
well known that for thermal light $ g^{(2)}=2 $. Therefore, we find enhanced
photon-bunching for our squeezed thermal light relative to thermal light.\par
Using Eqs.~(\ref{sinh_2}) and
(\ref{tanh_2u}), the second order correlation function in the steady state for arbitrary $g$ can be shown to be
\begin{equation}\label{g2_steady}
g^{(2)}_{ss}=2+\frac{4(n_{th}+\frac{1}{2})^2 (n_{th}^2 + n_{th})}{(\frac{2n_{th}}{g}\sqrt{n_{th}^2
+ n_{th}} +n_{th})^2}.
\end{equation}    
This quantity depends upon the thermal photon number and $g$, but not
upon the total photon number. Thus, to determine the steady state values, we need to know the number of thermal photons in the steady state as a function of $g$.

\begin{figure}          
                \includegraphics[scale=0.45]{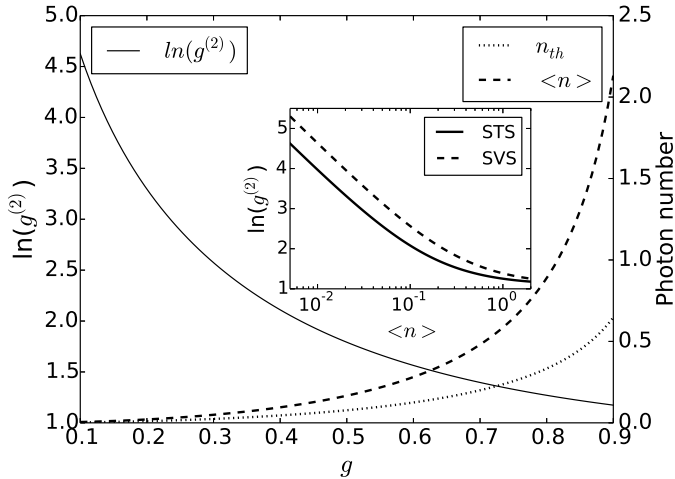}
           
                \caption{The second order correlation function (solid line), total photon number (dashed line) and thermal photon number (dotted line) as a function of pump-to-loss ratio, $g$. The inset shows the second order correlation function for a steady state squeezed thermal state generated with sub-critical pumping (solid) and a squeezed vacuum state (dashed) as a function of the total photon number.}
                \label{fig:analytic}
                \end{figure}

Using Eqs.(\ref{eq:b bdag}), (\ref{sinh_2}) and (\ref{tanh_2u}) one can show that in the weak pumping regime ($ g<1 $), the analytic expressions for the steady state total and thermal photon numbers are 
\begin{equation}\label{eq:n as g}
n_{th}=\frac{1-\sqrt{1-g^2}}{2\sqrt{1-g^2}}
\end{equation}
and
        \begin{equation}\label{eq:total n as g}
        \braket{n}=\frac{g^2}{2(1-g^2)}.
        \end{equation}
In Fig.~\ref{fig:analytic} we plot the thermal and total photon number as a function of $g$ in the steady state. As can be seen, both $ n_{th} $ and $ \braket{n}$ grow as we enhance the pumping power but at different rates. From Eqs.~(\ref{eq:n as g}) and (\ref{eq:total n as g}), it is easily shown that the fraction of thermal photons goes to 1/2 as $g$ goes to zero and that, although the number of photons diverges as $g \rightarrow 1$, the fraction of thermal photons goes to zero in the same limit.\par
From Eqs.~(\ref{g2_steady}) and (\ref{eq:n as g}), it can be shown that the photon anti-bunching decreases as we increase the pump power in the weak pumping regime. In other words, as shown in Fig.~\ref{fig:analytic}, the correlation function drops rapidly with $g$, reaching 3.2 at $ g=0.9 $, and finally reaching a value of 3 for $g=1$. We note that while there is a large $ g^{(2)} $ despite a small $ g $, this arises because there are very few photons in the cavity for these pump rates, resulting in photon antibunching. When the pumping is strong enough such that there is more than one photon in the cavity, $ g^{(2)}$ is only slightly larger than 3.
In order to compare the antibunching in the steady state for a leaky cavity to what one would obtain in a perfectly squeezed vacuum state with the same photon number, in the inset to Fig.~\ref{fig:g2_factor} we plot $ g^{(2)}$ as a function of the total photon number. For large $ \braket{n}$,  $ g^{(2)} \rightarrow 3$ for both the STS and the SVS, while for $ \braket{n} \rightarrow 0$, $ g^{(2)} $ for a SVS is twice that of a STS with same $ \braket{n} $, but both diverge. This shows that although the loss greatly affects the stretching in the $Y$ quadrature, it does not greatly affect the photon bunching except for very low pumping.  Therefore, in applications where the desirable property of the squeezed light is the antibunching, the cavity loss does not result in a significant degradation when $ \braket{n} \gtrsim 1 $. 

\subsection{Strong pumping regime}

\begin{figure}          
                \includegraphics[scale=0.45]{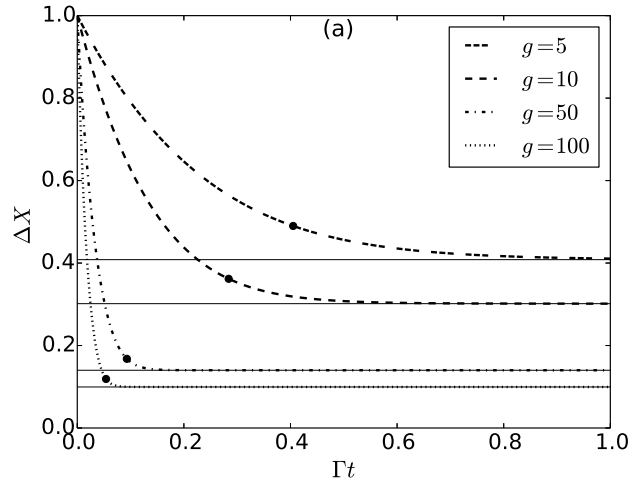}
                \includegraphics[scale=0.45]{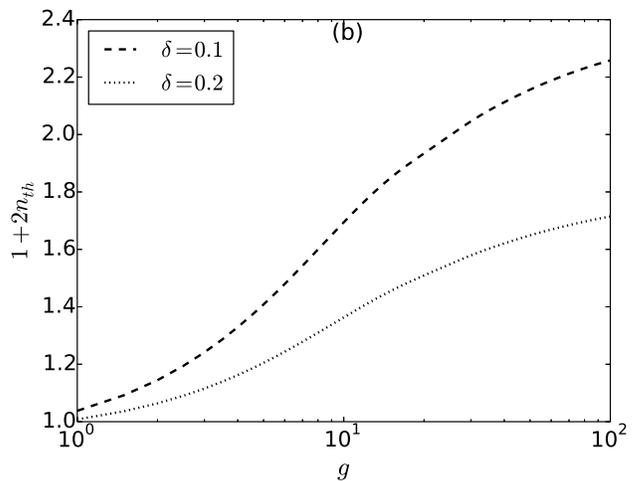}                
                \caption{(a) The quadrature noise in $X$ as a function of time in the strong pumping regime for $ g=5, 10, 50$ and $100 $. The steady state limits given by Eq.~(\ref{Xss}) are shown as solid lines, while the solid circles indicate the time when $ \Delta X $ reaches $ 20\% $ above the corresponding steady state value (i.e. $\delta=0.2$). (b)~The quantity $(2 n_{th} + 1)$ at the threshold times corresponding to $\delta= 0.1$ and $\delta = 0.2$ as a function of the pump-to-loss ratio, $ g $.}
                \label{fig:n_th vs g}
                \end{figure}

When one crosses into the strong pumping regime ($g>1$), there is no sudden change in the squeezing in $ X $, and by increasing $ g $ beyond unity the steady-state value of $ \Delta X $ can be decreased as much as desired. However, $ \Delta Y $ does not reach a steady-state value and instead increase in time without bound at an approximately exponential rate, causing a commensurate increase in $ \Delta X \Delta Y $ as the state evolves further away from an ideal squeezed state. In Fig.~\ref{fig:n_th vs g}(a), we plot $ \Delta X $ as a function of time for four different super-critical values of $g$. As expected, as $g$ increases, the steady-state squeezing in $ X $ is stronger and approaches the steady state limit of $ 1/\sqrt{1+g}$. It is clear that when the pump is stronger, the squeezing in $X$ more rapidly approaches its steady state value. To quantify this, we consider the quadrature squeezing at the time when $ \Delta X $ is just above the steady state limit such that  
\begin{equation}
\Delta X = \frac{1+\delta}{\sqrt{1+g}},
\end{equation}
where $ \delta $ is the percentages of deviation from the steady state limit. We refer to this as the threshold value. The small solid circles in Fig.~\ref{fig:n_th vs g}(a) indicate the times when $ \Delta X $ is $ 20\% $ above the corresponding steady state limit for each different value of $ g $. As can be seen, when $g$ is increased, this threshold value is reached much earlier. 

We now quantify the deviation from minimal uncertainty squeezing at the threshold point. According to Eqs.~(\ref{eq:delta x}) and (\ref{eq:delta y}), $\Delta X \Delta Y = 2n_{th}+1 $. Therefore, the quantity, $2n_{th}+1 $  can be used as a parameter to evaluate the deviation from minimum uncertainty squeezing. 

\begin{figure}          
                \includegraphics[scale=0.45]{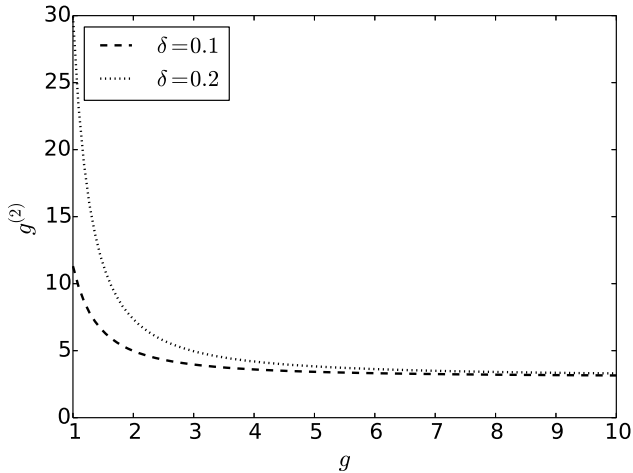}
            
                \caption{The second order correlation function at the threshold times corresponding to $\delta= 0.1$ and
$\delta = 0.2$ as a function of pump-to-loss ratio, $ g $.}
                \label{fig:smallg2VSg}
                \end{figure}

In Fig.~\ref{fig:n_th vs g} we plot $ (2n_{th}+1) $ at the $\delta = 0.1$ and $\delta = 0.2$ threshold values as a function of $g$.  As can be seen, the deviation from minimum uncertainty increases monotonically with $g$. However, even for a squeezing of 0.12 in $X$ ($g=100$, $\delta = 0.2$), the stretching in $Y$ is less than a factor of 2 above its minimal value of 8.4. Although this is certainly not negligible, it is less that the factor of 2 deviation that one obtains for a value of $g=\sqrt{3}/2 $ in the steady state, where there, the uncertainty in $X$ is only $0.732$. This shows that strong squeezing in a lossy cavity is best achieved by strongly pumping the system for a short time, not by working in the steady state or at long times. \\
We close this section by considering the second order correlation function in the strong pumping regime. In Fig.~\ref{fig:smallg2VSg} we plot $ g^{(2)}$ as a function of $g$ at the two different threshold values of $\Delta X$ corresponding to $\delta = 0.1$ and $\delta = 0.2$. As can be seen, for $g$ greater than about 5, $ g^{(2)}$ is close to 3. As $ g $ approaches unity $ g^{(2)} $ becomes large, but does not diverge. Note, however, that for the smaller values of $ g $ the time taken to reach steady state increases, and the number of photons in the cavity is rather small. For example, for $g = 1$ and $\delta = 0.1$, $\braket{n}=0.096$ and the threshold value is reached at $\Gamma t = 0.78$. 

\section{\label{sec:level6}Conclusion}
In this work, we have examined the properties of squeezed states generated in lossy cavities via degenerate parametric down conversion. We have solved the Lindblad master equation for a leaky cavity and have shown that the evolving state is a squeezed thermal state. \par
We have presented the calculated time dependence of the total and thermal photon numbers, the quadrature noise and the second order correlation function for different ratios of the pump power to the loss coefficient. Moreover, we have developed analytic expressions for the steady state values of these quantities in the weak pumping regime and have analysed their dependence on $g$ in the strong pumping regime. \par
We have found that although it is not possible to obtain steady-state squeezing that is greater than $1/\sqrt{2}$ when the pumping is sub-critical; for super-critical pumping, essentially arbitrarily strong squeezing is possible for large enough pump power.  Moreover, if the duration of the pump is kept sufficiently short, the deviation from minimal uncertainty can be kept relatively modest. Therefore if one wishes to obtain strong squeezing in a leaky cavity, as are all real cavities, it is necessary to operate in the strong pumping regime and to carefully time the duration and strength of the pump to achieve the desired squeezing in one quadrature, without obtaining excessive stretching in the other. \par
Finally, we have found that the squeezed thermal light that is obtained in steady state in the weak pumping regime results in strong antibunching that is comparable to that found is a pure squeezed vacuum state.  Similar antibunching can be achieved in the strong pumping at early times. Thus, such light could be used effectively in some quantum imaging techniques, such as ghost imaging~\cite{PhysRevA.93.013822}, where the large $ g^{(2)} $ results in a faithful image even when the source has low brightness.\par
\section{\label{sec:level7}ACKNOWLEDGMENTS}
We would like to thank Mohsen Kamandar Dezfouli for very enlightening
discussions and comments. We also thank the Natural Sciences and Engineering Research Council of Canada (NSERC) for financial support.
\appendix*\section{\label{sec:level7} Derivation of Eqs. (\ref{u}), (\ref{phi}) and (\ref{n})}  
In this Appendix, we supply some of the details of the derivation of the key differential equations, (\ref{u}), (\ref{phi})
and (\ref{n}). 
Using the chain rule, the time derivative of the $ O(t) $ operator can be written as
\begin{equation}\label{general_derivative}
\dot{O} (t)=\dot{O} _I(t)+\dot{O} _{II}(t)+\dot{O}_{III}(t)+\dot{O}_{IV}(t).
\end{equation}
We now evaluate each of the terms in Eq.~(\ref{general_derivative}) separately. 
The first term, $ \dot{O}_I $, is defined as
\begin{equation}
\begin{split}
\dot{O} _I  = & \frac{d\rho^{-1/2} _T}{dt}\rho^{1/2}_T\rho^{-1/2} _T S^{\dagger}\rho S \rho^{-1/2} _T  \\
& +\rho^{-1/2} _T S^{\dagger} \rho S \rho^{-1/2} _T \rho^{1/2} _T \frac{d\rho ^{-1/2} _T}{dt}, 
\end{split}
\end{equation}
which can be written as 
\begin{equation}\label{O_I}
\dot{O}_I =\{J,O\},
\end{equation} 
where 
\begin{equation}
J=\frac{d\rho^{-1/2} _T}{dt}\rho^{1/2} _T=\rho^{1/2} _T \frac{d\rho^{-1/2} _T}{dt}.
\end{equation}
Letting $ x=e^{-\beta \hbar \omega} $ we have
\begin{equation}
J=\frac{1}{2x}\frac{dx}{dt}(\frac{x}{1-x}-b^{\dagger} b).
\end{equation}
Using Eq.~(\ref{n and beta}), we obtain
\begin{equation} \label{eq:J}
J=\frac{1}{2x}\frac{dx}{dt}(n_{th}-b^{\dagger} b).
\end{equation}
The second term in Eq.~(\ref{general_derivative}) can be written as
\begin{equation}
\begin{split}
\dot{O}_{II} = & \rho ^{-1/2} _T \frac{dS^{\dagger}}{dt} S \rho ^{1/2} _T (\rho ^{-1/2} _T S^{\dagger} \rho S \rho ^{-1/2} _T)  \\
& +(\rho ^{-1/2} _T S^{\dagger} \rho S \rho ^{-1/2} _T) \rho ^{1/2} _T S^{\dagger} \frac{dS}{dt}  \rho ^{-1/2} _T, 
\end{split}
\end{equation}
which can be written as
\begin{equation}
\dot{O}_{II}=LO+OL^{\dagger},
\end{equation}
where 
\begin{equation}
L=\rho ^{-1/2}_T \frac{dS^{\dagger}}{dt} S \rho^{1/2} _T,
\end{equation}
\begin{equation}
L^{\dagger}=\rho ^{1/2}_T  S^{\dagger} \frac{dS}{dt} \rho^{-1/2} _T.
\end{equation}
The squeezing operator has the form of
\begin{equation}
S(\xi)=e^{\frac{1}{2}(\xi ^* b^2 - \xi b^{\dagger 2})},
\end{equation}
so we have
\begin{equation} \label{eq: L}
\begin{split}
L =& (-i s^2 \dot{\phi}) (b^{\dagger}b+\frac{1}{2})\\
& +\frac{1}{2}\dot{u} (x^{-1} b^{\dagger 2} e^{i \phi}- x b^{2} e^{-i \phi}) \\
& + \frac{1}{2} ics \dot{\phi} (x^{-1} b^{\dagger 2} e^{i \phi}+ x b^{2} e^{-i \phi}),
\end{split}
\end{equation}
where $ s \equiv \sinh u $ and $ c \equiv \cosh u $. Noting that $ L $ can be written as $L=M+iN $, we can finally write $ \dot{O}_{II} $ as
\begin{equation}\label{O_II}
\begin{aligned}
\dot{O}_{II} &= (M+iN)O+O(M-iN) \\
&= \{M,O\}+i[N,O].
\end{aligned}
\end{equation}
We write the third term in Eq.~(\ref{general_derivative}) as
\begin{equation}\label{O_III}
\dot{O}_{III}=\dot{O}_{0}+\dot{O}_{V}+\dot{O}_{L},
\end{equation} 
where the first two terms are from the Hamiltonian evolution, and the second term is from the Lindblad contribution. In particular,
\begin{equation}\label{O_0}
\dot{O}_0 = \rho^{-1/2} _T S^{\dagger}(-\frac{i}{\hbar}[\hbar \omega b^{\dagger}b,\rho])S\rho^{-1/2}_T
\end{equation} 
and
\begin{equation}\label{O_V}
\dot{O}_V = \rho^{-1/2} _T S^{\dagger}(-\frac{i}{\hbar}[V(t),\rho])S\rho^{-1/2}_T
\end{equation}
and of course
\begin{equation}\label{O_L}
\begin{split}
\frac{\dot{O}_L}{\Gamma} &=\rho^{-1/2}_T S^{\dagger} b \rho b^{\dagger} S \rho ^{-1/2} _T \\
& -\frac{1}{2}\rho^{-1/2} _T S^{\dagger}(-\frac{i}{\hbar}\{b^{\dagger}b,\rho\})S\rho^{-1/2}_T. 
\end{split}
\end{equation} 
From Eq.~(\ref{O_0}) we have
\begin{equation}
\frac{\dot{O}_0}{-i\omega}=GO-OG^{\dagger},
\end{equation}
where 
\begin{equation}\label{G}
G=\rho^{-1/2}_T S^{\dagger} b^{\dagger} S \rho^{1/2}_T.
\end{equation}
Writing $ G $ as $ G=P+iQ $, where both $ P $ and $ Q $ are Hermitian, we have
\begin{equation}\label{O_01}
\dot{O}_0=-i\omega [P,O]+\omega \{Q,O\}.
\end{equation}
Following the same method for Eq.~(\ref{O_V}) and Eq.~(\ref{O_L}) we have
\begin{equation}\label{O_V1}
\dot{O}_V (t)=-\frac{i}{\hbar}[\bar{P},O]+\frac{1}{\hbar}\{\bar{Q},O\}
\end{equation}
and
\begin{equation}\label{O_L1}
\dot{O}_L=\Gamma F - \frac{1}{2} \Gamma \{P,O\}-\frac{i}{2}\Gamma [Q,O],
\end{equation} 
where 
\begin{equation}
F=\rho^{-1/2}_T S^{\dagger} b \rho b^{\dagger} S \rho ^{-1/2} _T = TOT^{\dagger}
\end{equation}
\begin{equation}\label{eq:T}
T=\rho^{-1/2} _T S^{\dagger}b \rho S \rho^{1/2} _T
\end{equation}
\begin{equation}
\begin{split}
\bar{P}&=-cs(\gamma \alpha e^{-i \phi}+\gamma ^{*} \alpha^ {*} e^{i \phi})\\
&-2cs(\gamma \alpha e^{-i \phi}+\gamma ^{*} \alpha^ {*} e^{i \phi})b^{\dagger}b\\
&+\frac{1}{2}(\gamma\alpha(x^{-1}+x)c^{2}+\gamma^{*}\alpha^{*}(x^{-1}+x)s^{2}e^{2i\phi})b^{\dagger}b^{\dagger}\\
&+\frac{1}{2}(\gamma\alpha(x^{-1}+x)s^{2}e^{-i\phi}+\gamma^{*}\alpha^{*}(x^{-1}+x)c^{2})bb
\end{split}
\end{equation}
\begin{equation}\label{Qbar}
\begin{split}
\bar{Q} &=-\frac{i}{2}(x^{-1}-x)(\gamma^{*}\alpha^{*}s^{2}e^{2i \phi}+\gamma \alpha c^{2})b^{\dagger}b^{\dagger}\\
& +\frac{i}{2}(x^{-1}-x)(\gamma\alpha s^{2}e^{-2i \phi}+\gamma^{*} \alpha^{*} c^{2})bb.
\end{split}
\end{equation}
\\

We can now assemble the dynamic equation for $ O(t) $. From Eqs.~(\ref{O_I}, \ref{O_II}, \ref{O_III}, \ref{O_01}, \ref{O_V1}, \ref{O_L1}) we have
\begin{equation}
\begin{split}
\dot{O}(t)&=\{J+M+\omega Q+\frac{1}{\hbar}\bar{Q},O\}-i[\omega P+\frac{1}{\hbar}\bar{P}-N,O]\\
&+\Gamma TOT^{\dagger}-\frac{1}{2}\Gamma\{P,O\}-\frac{i}{2}[Q,O].
\end{split}
\end{equation}
Initially $ O(t=0) = I$. If we want $ O(t) $ to remain the identity operator for all time, then the right hand side of this equation must vanish if $ O $ is replaced by $ I $. The commutators clearly vanish, and we see that the condition that must be satisfied is
\begin{equation} \label{O_finall}
2J+2M+2\omega Q+\frac{2}{\hbar}\bar{Q}+\Gamma (TT^{\dagger}-P)=0.
\end{equation}
Using Eq.~(\ref{eq: L}) and considering $ L=M+iN $ we can write
\begin{equation}\label{eq:M}
\begin{split}
M &= \frac{1}{2} (L+L^{\dagger})\\
&= \frac{1}{4} \dot{u} (x^{-1}-x)((b^{\dagger})^{2} e^{i \phi} + b^{2} e^{-i \phi} )\\
&+ \frac{1}{4} ics\dot{\phi}(x^{-1} - x)((b^{\dagger})^{2} e^{i \phi} - b^{2} e^{-i \phi} )
\end{split}
\end{equation}
and
\begin{equation}
\begin{split}
N &= \frac{L-L^{\dagger}}{2i}\\
&=-s^{2} \dot{\phi}(b^{\dagger} b +\frac{1}{2}) \\
&- \frac{1}{4} i\dot{u} (x^{-1}+x)((b^{\dagger})^{2} e^{i \phi} - b^{2} e^{-i \phi} )\\
&+ \frac{1}{4} cs\dot{\phi}(x^{-1} + x)((b^{\dagger})^{2} e^{i \phi} + b^{2} e^{-i \phi}).
\end{split}
\end{equation}
By considering $ G=P+iQ $ and expanding Eq.~(\ref{G}) as
\begin{equation}
\begin{split}
G=s^2+(c^2+s^2)b^{\dagger}b-cs(x^{-1}(b^{\dagger})^{2} e^{i \phi}+x(b)^2 e^{-i \phi}),
\end{split}
\end{equation}
we have
\begin{equation}\label{eq:P}
\begin{split}
P &= \frac{G+G^{\dagger}}{2}
 =s^2 +(c^2+s^2)b^{\dagger}b\\
&-\frac{1}{2}cs(x^{-1}+x)((b^{\dagger})^{2} e^{i \phi}+x(b)^2 e^{-i \phi})
\end{split}
\end{equation}
and
\begin{equation} \label{eq:Q}
\begin{split}
Q &= \frac{G-G^{\dagger}}{2i}\\
&=\frac{1}{2}ics(x^{-1}-x)((b^{\dagger})^{2} e^{i \phi}-(b)^2 e^{-i \phi}).
\end{split}
\end{equation}
Eq.~(\ref{eq:T}) can also be expanded as
\begin{equation}
T=x^{1/2}cb-x^{-1/2}se^{i \phi}b^{\dagger},
\end{equation}
which allows us to write $ TT^{\dagger} $ as
\begin{equation}\label{eq:TT}
TT^{\dagger}=xc^2+(xc^2+x^{-1}s^2)b^{\dagger}b -cs(b^2 e^{-i \phi}+(b^{\dagger})^{2} e^{i \phi}).
\end{equation}
Substituting $ J, M, Q, \bar{Q}, TT^{\dagger}, P  $ in Eq.~(\ref{O_finall}) with Eqs.~(\ref{eq:J}, \ref{eq:M}, \ref{eq:Q}, \ref{Qbar}, \ref{eq:P}, \ref{eq:TT}), we have
\begin{equation}\label{eq: sipe43}
\begin{split}
0 &= \frac{1}{x}\frac{dx}{dt}(n_{th}-b^{\dagger}b)\\
&+ \frac{1}{2}\dot{u}(x^{-1}-x)((b^{\dagger})^{2} e^{i \phi}+(b)^2 e^{-i \phi})\\
&+ \frac{1}{2}ics\dot{\phi}((b^{\dagger})^{2} e^{i \phi}-(b)^2 e^{-i \phi})\\
&+i \omega cs(x^{-1}-x)((b^{\dagger})^{2} e^{i \phi}-(b)^2 e^{-i \phi})\\
&+\frac{i}{\hbar}(x^{-1}-x)(\gamma \alpha s^2 e^{-2i \phi} + \gamma^{*} \alpha^{*} c^2)bb\\
&-\frac{i}{\hbar}(x^{-1}-x)(\gamma^{*} \alpha^{*} s^2 e^{2i \phi} + \gamma \alpha c^2)b^{\dagger}b^{\dagger}\\
&+\Gamma(xc^2 + (xc^2 + x^{-1}s^{2})b^{\dagger}b-cs(b^2 e^{-i \phi} + (b^{\dagger})^2 e^{i\phi}))\\
&-\Gamma(s^2+(c^2+s^2)b^{\dagger}b-\frac{1}{2}cs(x^{-1}+x)((b^{\dagger})^{2} e^{i\phi}+b^2 e^{-i \phi})).
\end{split}
\end{equation} 
To simplify this a bit, we introduce two Hermitian operators
\begin{equation}
\chi_1=(b^{\dagger})^2 e^{i\phi}+b^2 e^{-i \phi},
\end{equation}
\begin{equation}
\chi_2=i((b^{\dagger})^2 e^{i\phi}-b^2 e^{-i \phi}).
\end{equation}
Then
\begin{equation}
\chi_1 - i\chi_2 = 2(b^{\dagger})^2 e^{i \phi},
\end{equation}
so that
\begin{equation}
(b^{\dagger})^2 = \frac{1}{2} e^{-i \phi}(\chi_1 - i\chi_2),
\end{equation}
\begin{equation}
b^2 = \frac{1}{2} e^{i \phi}(\chi_1 + i\chi_2)
\end{equation}
and in terms of these we can write Eq.~(\ref{eq: sipe43}) as
\begin{equation}
\begin{split}
0 &= \frac{1}{x}\frac{dx}{dt}(n_{th}-b^{\dagger}b)\\
&+ \frac{1}{2}\dot{u}(x^{-1}-x)\chi_1 + \frac{1}{2}cs\dot{\phi}(x^{-1}-x)\chi_2\\
&+ \omega cs (x^{-1}-x)\chi_2\\
&+\frac{i}{2\hbar}(x^{-1}-x)(\gamma \alpha s^2 e^{-i \phi} + \gamma^{*} \alpha^{*} c^{2} e^{i \phi})(\chi_1 +i\chi_2)\\
&-\frac{i}{2\hbar}(x^{-1}-x)(\gamma^{*} \alpha^{*} s^2 e^{i \phi} + \gamma \alpha c^{2} e^{-i \phi})(\chi_1 -i\chi_2)\\
&+ \Gamma (xc^2 + (xc^2 + x^{-1}s^2)b^{\dagger}b-cs\chi_1)\\
&- \Gamma (s^2+(c^2+s^2)b^{\dagger}b-\frac{1}{2}cs(x^{-1}+x)\chi_1)
\end{split}
\end{equation}
or
\begin{equation}
0=F_1 \chi_1 + F_2 \chi_2 + F_3 b^{\dagger}b+F_4,
\end{equation}
where
\begin{equation}
\begin{split}\label{eq:F_1}
F_1 &= \frac{1}{2}\dot{u} (x^{-1}-x)\\
&+\frac{i}{2\hbar}(x^{-1}-x)(\gamma^{*} \alpha^{*} e^{i \phi} - \gamma \alpha  e^{-i \phi})\\
&-\Gamma cs+\frac{1}{2}\Gamma cs(x^{-1}+x),
\end{split}
\end{equation}
where we have used the fact that $ c^2 -s^2 =1 $,
\begin{equation}
\begin{split}\label{eq:F_2}
F_2 &=\frac{1}{2}cs \dot{\phi}(x^{-1}-x)+\omega cs (x^{-1}-x)\\
&- \frac{1}{2\hbar} (x^{-1}-x)(c^2 + s^2)(\gamma \alpha e^{-i \phi} + \gamma^{*} \alpha^{*} e^{i \phi})
\end{split}
\end{equation}
and
\begin{equation}\label{eq:F_3}
F_3=-\frac{1}{x}\frac{dx}{dt}+\Gamma (xc^2 + x^{-1} s^2)-\Gamma (c^2 + s^2)
\end{equation}
and finally
\begin{equation}\label{eq:F_4}
F_4=\frac{1}{x}\frac{dx}{dt}n_{th}+\Gamma (xc^2-s^2).
\end{equation}
To have a solution we must have
\begin{equation}\label{eq:F=0}
F_1 = F_2 =F_3 =F_4 =0
\end{equation}
From Eq.~(\ref{eq:F_1}) and Eq.~(\ref{eq:F=0}) we have
\begin{equation}
\begin{split}
\dot{u}&=\frac{i}{\hbar}(\gamma \alpha e^{-i \phi} - \gamma^{*} \alpha^{*} e^{i \phi})+\frac{2\Gamma cs}{(x^{-1}-x)} \\
&- \Gamma cs \frac{(x^{-1}+x)}{(x^{-1}-x)},
\end{split}
\end{equation}
which can be written in the form of Eq.~(\ref{u}) as
\begin{equation}\label{eq:sipe51}
\dot{u}=\frac{i}{\hbar}(\gamma \alpha e^{-i \phi} - \gamma^{*} \alpha^{*} e^{i \phi})-\frac{\Gamma cs}{2n_{th}+1}.
\end{equation}
Next, from Eq.~(\ref{eq:F_2}) and Eq.~(\ref{eq:F=0}) we have
\begin{equation}
\frac{1}{2}cs \dot{\phi}+\omega cs -\frac{1}{2\hbar}(c^2 + s^2)(\gamma \alpha e^{-i \phi} + \gamma^{*} \alpha^{*} e^{i \phi})=0
\end{equation}
or
\begin{equation}\label{eq:sipe52}
\dot{\phi}=-2\omega+\frac{1}{\hbar}\frac{c^2 + s^2}{cs}(\gamma \alpha e^{-i \phi} + \gamma^{*} \alpha^{*} e^{i \phi}),
\end{equation}
Which is in agreement with Eq.~(\ref{phi}).
Turning next to Eq.~(\ref{eq:F_3}) and Eq.~(\ref{eq:F=0}) we have
\begin{equation}\label{eq:sipe53}
\frac{1}{x}\frac{dx}{dt}=\Gamma (xc^2+x^{-1}s^2 -c^2 -s^2),
\end{equation}
or
\begin{equation}
\frac{dx}{dt}=\Gamma(1-x)(s^2 (1-x)-x).
\end{equation}
Now from Eq.~(\ref{n and beta}) and reminding that $ x=e^{-\beta \hbar \omega} $, we have
\begin{equation}
\begin{split}
\frac{dn_{th}}{dt}&=\frac{d}{dt}(\frac{x}{1-x})\\
&= \frac{1}{(1-x)^2}\frac{dx}{dt},
\end{split}
\end{equation}
so from Eq.~(\ref{eq:sipe53}) we have
\begin{equation}
\begin{split}
\frac{dn_{th}}{dt}&=\Gamma(s^2-\frac{x}{1-x})\\
&= \Gamma (s^2 - n_{th}),
\end{split}
\end{equation}
which is in agreement with Eq.~(\ref{n}).\\

%
%

\begin{thebibliography}{41}
\bibitem [{ext()}]{extra1}%
  \BibitemOpen
  \href@noop {} {}\bibinfo {note} {We note that when Eqs.~(\ref{n}) and
  (\ref{modified_u}) are also valid for $g\rightarrow g(t) $, where $g(t)$ is a
  real function of time.}\BibitemShut {Stop}%
\end{thebibliography}
\end{document}